\documentclass[11pt,twoside]{article}

\usepackage{asp2006}
\usepackage{fancyhdr,graphicx,caption,rotating,multirow,lscape}
\DeclareGraphicsExtensions{.eps,.eps.gz,.ps,.jpg,.tiff,.bb}
\begin{document}

%%% Fill in title
\title{Developments in Planet Detection using Transit Timing Variations}

%%% Fill in author names, use initials and surname and affiliation
%%% One author
\author{J. H. Steffen}
\affil{Fermi National Accelerator Laboratory, Batavia, IL 60510 USA}

%%% Two authors and more
\author{E. Agol}
\affil{University of Washington, Department of Astronomy, Seattle, WA 98195 USA}

\begin{abstract} %%% Abstract to run on from here.
In a transiting planetary system, the presence of a second planet will cause the time interval between transits to vary.  These transit timing variations (TTV) are particularly large near mean-motion resonances and can be used to infer the orbital elements of planets with masses that are too small to detect by any other means.  I present the results of a study of simulated data where I show the potential that this planet detection technique has to detect and characterize secondary planets in transiting systems.  These results have important ramifications for planetary transit searches since each transiting system presents an opportunity for additional discoveries through a TTV analysis.  I present such an analysis for 13 transits of the HD 209458 system that were observed with the Hubble Space Telescope.  This analysis indicates that a putative companion in a low-order, mean-motion resonance can be no larger than the mass of the Earth and constitutes, to date, the most sensitive probe for extrasolar planets that orbit main sequence stars.  The presence or absence of small planets in low-order, mean-motion resonances has implications for theories of the formation and evolution of planetary systems.  Since TTV is most sensitive in these regimes, it should prove a valuable tool not only for the detection of additional planets in transiting systems, but also as a way to determine the dominant mechanisms of planet formation and the evolution of planetary systems.
\end{abstract}

%%% MAIN BODY OF TEXT GOES HERE. 
%%% Invited Talks : 10 pages
%%% Talks         : 6  pages
%%% Posters       : 3  pages
%%%IF YOU FEEL THAT THE INFORMATION PROVIDED 
%%% IN THIS FILE IS NOT SUFFICIENT, CONSULT THE FILE aspauthor2006.pdf. IT
%%% CONTAINS "INSTRUCTIONS FOR AUTHORS USING LATEX2E MARKUP", 
%%% SECTIONS 2.3-2.6 FOR HELP WITH EQUATIONS, FIGURES, AND TABLES.

\section{Introduction}

The core accretion model of planet formation often predicts the presence of small planet masses that become trapped in mean-motion resonance with a gas giant planet as the giant planet migrates inward \citep{tho05,zho05,pap05,ray06}.  The small sizes and masses of these putative trapped planets render them difficult to detect by conventional methods like radial velocity (RV) measurements and planetary transits.  However, recent work by \cite{ago05} and \cite{hol05} indicate that, in a known transiting system, the times of the planetary transits are a particularly sensitive probe for additional, resonant bodies.

If a second planet exists in a transiting system, then dynamical interactions within the system cause the time interval between transits to vary.  For systems in mean-motion resonance, these transit timing variations (TTV) can be quite large; often of order $(m_{\rm p}/m_{\rm T}) P_{\rm T}$ in where the $m_{\rm p}$ and $m_{\rm T}$ are the masses of the perturbing and transiting planets respectively and $P_{\rm T}$ is the period of the transiting planet (this amounts to nearly 15 minutes for an Earth-mass perturber and a Jupiter-mass transiting planet that is in a 3-day orbit)\footnote{We note that \cite{ago05} and \cite{hol05} define the TTV signal differently.  The former being the residuals from a linear fit to the period and epoch, the latter being the difference between successive periods.  The former gives a larger signal for resonant systems while for non-resonant systems the signals are comparable.  When analyzing data, one will likely fit the transit times directly and will thus render this difference immaterial.}.  From this TTV signal, one may infer the orbital elements and mass of the perturbing planet.

The strong sensitivity of TTV to resonant systems should allow for the discovery of sub Earth-mass planets in transiting systems, possibly using modest ($\sim 1$ m) ground-based telescopes.  Indeed, the first TTV analysis was conducted for the TrES-1 system where ground-based observations of 11 transits were able to probe for planets with masses smaller than the Earth \citep{ste05}.  Here we present portions of a similar study of Hubble Space Telescope (\textit{HST}) observations of 13 transits of the HD 209458 system \citep{bro01,sch04,knu06}.  We compare the sensitivity to mass of ground-based observations with space-based observations.  We also present the results of an analysis of 100 simulated systems where we attempt to infer the orbital elements of each system from a set of transit times.

\section{Analysis of HD 209458 System}

\begin{figure}[!t]
\centering
\includegraphics[angle=0,width=9cm]{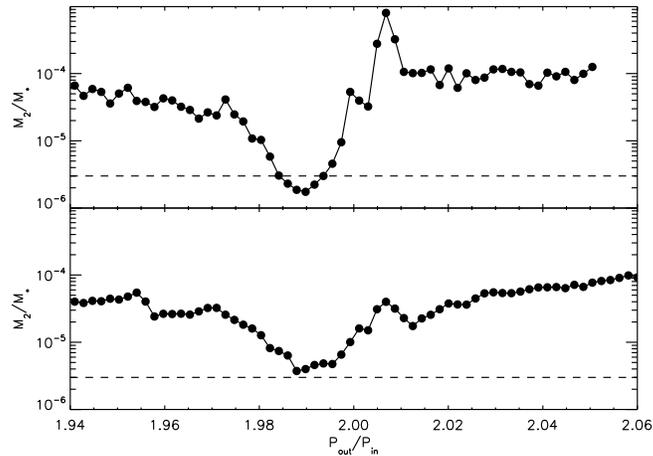}
\caption{Maximum allowed mass of a secondary planet in the HD 209458 system near the 2:1 interior (upper panel) and exterior (lower panel) resonances for any value of eccentricity.  The dashed lines correspond to the mass of the Earth.\label{zoom21}}
\end{figure}

We analyzed 13 transit times that were obtained from observations of HD 209458 using various instruments on \textit{HST}.  The details of the photometric data reduction, the transit times, and the balance of our results appear in \cite{ago06}.  While those data give no evidence for additional planets, we use them to constrain the presence of secondary planets as a function of various orbital elements.  Of particular interest are the constraints on planets in the 2:1 mean-motion resonances (see Figure \ref{zoom21}).  We find that a secondary planet in the exterior resonance cannot be much more massive than the Earth.  For the interior resonance, a second planet must be significantly less massive than the Earth in order for the systems to be consistent with the data at the 3-$\sigma$ level.  

\begin{figure}[!t]
\centering
\includegraphics[angle=0,width=9cm]{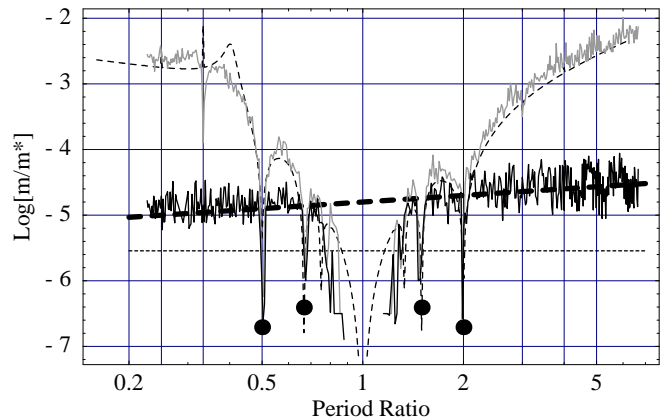}
\caption{Maximum allowed mass for a second planet on an initially circular orbit in the HD 209458 system as a function of its period ratio with the known planet.  The gray curve is the result from a TTV analysis of the transit times from \cite{ago06}, the dotted curve is the theoretical prediction from the appendix of \cite{ago05}, the black curve is the result of simultaneously analyzing the transit times and the RV measurements of \cite{lau05}, the thick dashed line is the prediction for the sensitivity of RV measurements from \cite{ste05}, the large dots are given by equation 33 of \cite{ago05}, and the horizontal dashed line corresponds to the mass of the Earth.\label{massivegraph}}
\end{figure}

In addition, we simultaneously analyzed these transit data and 68 of the RV measurements reported by \cite{lau05} in effort to constrain the allowed mass of a second planet.  Here, we restrict our analysis to initially circular orbits.  Figure \ref{massivegraph} shows the results of this analysis and compares it with the theoretical predictions given in \cite{ago05} and \cite{ste05}.  We find that for many orbits, particularly those where the period ratio is larger than 2:1, the RV technique is more sensitive than TTV.  It also illustrates the tremendous advantage that TTV has in probing resonant systems.

Our TTV analysis cannot exclude the putative secondary planet that was proposed by \cite{bod03} as an explanation for the apparently large radius of HD 209458b.  The proposed planet is relatively distant (having a period of $\sim 80$ days) and is not near a mean-motion resonance.  Consequently, it would induce a TTV signal with an amplitude less than one second.  RV measurements are better suited to test this hypothesis which we did not attempt.

\section{Ground-based vs. Space-based}

The transit times for the TrES-1 system that were published by \cite{cha05} have an average timing error of about 100 seconds.  The most precise of the times, derived from observations with the 1.2m Fred Whipple telescope, had an uncertainty of 26 seconds.  In contrast, the average uncertainty for the \textit{HST} observations of HD 209458 is 25 seconds with the best being 10 seconds.  This serves to illustrate the opportunities available to ground-based facilities.  If the timing precision of 26 seconds is reproducible, then a set of such observations could readily probe for resonant planets with masses smaller than the Earth.

Recent studies by \cite{hol06} indicate that $\sim 10$ second timing precision should be possible with ground-based telescopes.  This fact, and the relative ease of obtaining time on such telescopes compared with their space-based counterparts, shows that ground-based, follow-up observations of transiting systems identified by \textit{Kepler} and \textit{CoRoT} should prove invaluable.  Those observations could address important issues in planet formation and evolution and may identify the smallest known extrasolar planets for the foreseeable future.

\section{Identifying Planets with TTV}

Once a TTV signal is identified there remains the question of how difficult it is to infer the orbital elements of the perturbing planet.  We outline here our preliminary investigation of this issue.  For this study we randomly generated 100 systems that are near the 2:1 mean-motion resonance where the perturbing planet is exterior to the transiting planet.  The distributions from which we drew our orbital elements are uniform for the longitude of pericenter, time of pericenter passage, and period ratio (within 5\% of the resonance) and logarithmically uniform in eccentricity and mass.  For the unknown planet, the eccentricities are between $10^{-3}$ and 1 and the masses are between $10^{-6}M_0$ and $10^{-3}M_0$ where $M_0$ is the mass of the central star.  The mass of the known planet was fixed at $10^{-3}M_0$ and its eccentricities are between $10^{-3.5}$ and $10^{-0.3}$.  We rejected any system where the hill spheres of the two planets could touch.  We chose resonant systems because that is the regime where the TTV technique is unique in its sensitivity and where it will likely prove most valuable.  We allow the perturbing planet masses to be so large (large enough to be seen with RV measurements) because it may be possible to conduct a TTV analysis of \textit{Kepler} and \textit{CoRoT} data before RV measurements are available.

For each system we generate a set of 120 consecutive transit times.  To each time we add Gaussian, white timing noise with a dispersion equal to 5 seconds per day of the period of the transiting planet (e.g. for a 3-day period transiting planet the timing error is 15 seconds).  From these transit times we analyze 5 subsets of data: (1) all 120 transit times, (2) 60 randomly selected times, (3) 30 random times, (4) 15 random times, and (5) 15 correlated times where, if a particular transit is selected then there is a 67\% chance that a second transit is selected within 4 orbits and, if a second transit is selected, then there is a 50\% chance that a third transit is selected within 4 orbits of the second transit.  This correlated selection should roughly approximate the effects of realistic programs which are constrained by observation conditions, seasonal variations, etc. though these criteria were not selected with any serious rigor.

For our analysis, we assumed that the orbital elements of the transiting planet are known and that the orbits of the two planets are coplanar.  We systematically step through the period ratio of the planets between the 3:2 and the 7:1 resonances and then marginalize over the eccentricity, longitude of pericenter, time of pericenter passage, and mass of the perturber at each point.  This marginalization is done by choosing 500 random values for the parameters at each value of the period ratio, generating hypothetical transit times for each system, calculating the $\chi^2$ between the actual transit times and the hypothetical transit times, and requiring the set of parameters with the lowest $\chi^2$ to represent the best-fit system for that value of the period ratio.

\begin{figure}[!t]
\centering
\includegraphics[angle=0,width=9cm]{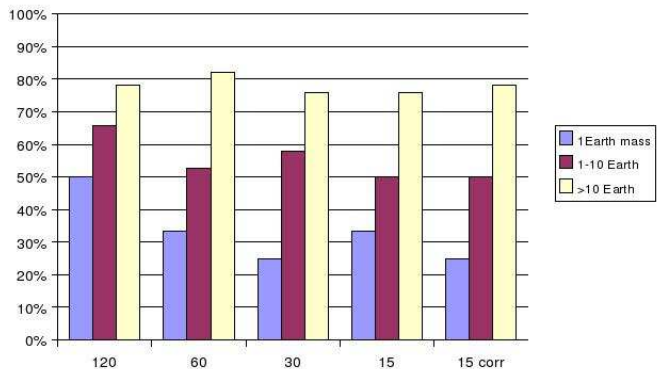}
\caption{The fraction of correctly identified planets in the exterior 2:1 mean-motion resonance with a known planet for analyzes of 120, 60, 30, 15, and 15 correlated transit observations.  The bar on the left of each set is for planets with masses less than the mass of the Earth.  The center bar of each set is for planets with masses between one and ten Earth masses.  The bar on the right is for planets larger than 10 times the mass of the Earth.\label{results}}
\end{figure}

For each system and each set of transit timing data we consider the identification of the perturbing planet to be successful if one of the three lowest values of the $\chi^2$---that is, the $\chi^2$ as a function of period ratio---is within 5\% of the actual period ratio.  Often the remaining orbital elements are also correctly identified, but for our purposes we neglect this information since, once the period ratio is identified a more detailed analysis could be done to find the remaining parameters and their uncertainties.  Our results are shown in Figure \ref{results} where we see that a significant fraction of the systems are correctly identified.

These preliminary results likely represent a lower bound for the fraction of systems that can be identified from such data.  A more sophisticated search algorithm will certainly do better than the one that we use, which is chosen for simplicity and speed and whose scope is only to provide some basic insight into the challenges associated with inferring the parameters.  Some common characteristics of the systems which we fail to identify are that the perturber masses are often small, the eccentricities are often large, and the period ratios that have the lowest $\chi^2$ are correspond to resonances other than the 2:1 resonance.  Such resonances often have similar TTV signals over the time span of the data.

We mention one issue exposed by this experiment.  For a TTV analysis of transit data it is unlikely that a basic Markov Chain Monte Carlo (MCMC) algorithm will be appropriate.  This is because, for a system in resonance (where TTV is most useful) the $\chi^2$ often has many deep, narrow local minima that correspond to various resonances.  Such minima are difficult for a Markov Chain to handle because, if the step size of the chain is too small it will become trapped in these local minima.  Yet, if it is large enough to escape the local minima then it is often too large to properly resolve them.  Thus, a multi-stage investigation will probably prove more fruitful where, for example, the local minima are first identified using a search similar to the one outlined above and then each minima is studied with MCMC such that the chain remains inside the selected minima.

\section{Discussion}

The results presented in this article indicate that TTV should be a valuable tool for both discovering additional planets and for constraining the relative importance of different planet formation and evolution mechanisms.  A set of a few tens of high quality, ground-based transit observations can readily probe for Earth-mass or smaller planets in several mean-motion resonances.  A similar number of transit observations with space-based telescopes should be able to identify mars-mass companions.  Such discoveries could place constraints on the initial mass function of the planetesimals and would provide important guidance to theorists who model planet formation.  We showed that relatively simple search algorithms can correctly identify a large fraction of possible resonant systems.  Thus, if small resonant bodies do exist in planetary systems we are confident that they can be found using the TTV technique.

\acknowledgements %%% Text of acknowledgments runs on after this command.
Jason Steffen is supported by the Brinson Postdoctoral Fellowship.

%%% THE BIBLIOGRAPHY

\end{document}